\documentclass[twocolumn,showpacs,preprintnumbers,amsmath,amssymb]{revtex4}

\usepackage{graphicx}
\usepackage{dcolumn}
\usepackage{bm}

\begin{document}

\title{Elastic and magnetic properties on field-induced Metal-Insulator transition\\ of bilayer manganese oxide (La$_{1-z}$Pr$_{z}$)$_{1.2}$Sr$_{1.8}$Mn$_{2}$O$_{7}$ ($z$=0.6)}

\author{Y. Nakanishi}
\email{yoshiki@iwate-u.ac.jp}
\author{K. Shimomura}
\author{T. Kumagai }
\author{M. Matsukawa}
\author{M. Yoshizawa}

\affiliation{%
Graduate School of Frontier Materials Function Engineering, Iwate University Morioka 020-8551, Japan
}%

\author{M. Apostu}
\author{R. Suryanarayanan }
 \author{A. Revcolevschi}
\affiliation{Laboratoire de Physico-Chimie de L'etat Solide, CNRS, UMR 8648 Universit\'{e} Paris-Sud, 91405 Orsay, France 
}%

\author{S. Nakamura}
\affiliation{ Center for Low-temperature Science, Tohoku University, Sendai 980-8578, Japan 
}%

\date{\today}

\begin{abstract}
The elastic and magnetic properties of a single crystal of 327 bilayer manganese oxide (La$_{1-z}$Pr$_{z}$)$_{1.2}$Sr$_{1.8}$Mn$_{2}$O$_{7}$ with $z$=0.6 have been investigated by ultrasonic and high-field magnetization measurements. A distinct elastic anomaly was observed at low temperatures and in high magnetic fields when crossing the phase boundary. A pronounced elastic softening as a function of magnetic field ($H$) appears across the boundary of the low-temperature magnetic phase below around 40 K, accompanied by a distinct hysteresis. In the high field region, however, these elastic constants exhibit an increase with increasing a magnetic field in the high-field region. The high-field magnetization measurement characterizing the magnetic state clarified a strong connection between elastic strains and magnetic moments. The new metamagnetic transition was found in the high-field region above 40 K, accompanied with no hysteresis. The present results of elastic constants are discussed in terms of a coupling effect between elastic strains associated with sound waves and relevant magnetic moments, and relevant conduction bands. They are explained reasonably in terms of a strong coupling between elastic strains and magnetic susceptibility $\chi$$_{m}$=$\partial$$M$/$\partial$$H$. 

\end{abstract}

\maketitle

\section{\label{sec:level1}Introduction}

The interesting and anomalous properties of the manganites showing a colossal magnetoresistance (CMR) compounds are thought to originate from the competition between charge, lattice and spin degrees of freedom. The fundamental mechanism of CMR is reasonably understood within a framework of the double exchange model, which is based on the hopping effect of $e$$_{g}$ electrons between Mn$^{3+}$ and Mn$^{4+}$ ions [1-2]. In such a case the ferromagnetic interaction arises from the strong on-site coupling between the charge carriers ($e$$_{g}$-state) and the local spins ($t$$_{2g}$-state). However, there are many experimental reports that cannot be explained by this framework at present. Since the discovery of the larger magnitude of CMR in so-called 327 bilayer manganese oxide: La$_{2-2x}$Sr$_{1+2x}$Mn$_{2}$O$_{7}$, a considerable research effort has been devoted to the understanding of the properties of these compounds [3-4]. Furthermore, the substitution of Pr$^{3+}$ for La ions at the concentration of $z$=0.6 in (La$_{1-z}$Pr$_{z}$)$_{1.2}$Sr$_{1.8}$Mn$_{2}$O$_{7}$ (hereafter denoted as LPSMO) realizes the system with a record value of the CMR in this bilayer manganese oxide family, which amounts to a huge decrease of the $c$-axis resistivity by a factor of one million in the presence of a magnetic field [5-6]. The first-order transition is accompanied by a remarkable negative magnetostriction, indicating the presence of important interplay among spin, carrier and orbital degrees of freedom [7-8].

The host compound, La$_{1.2}$Sr$_{1.8}$Mn$_{2}$O$_{7}$ exhibits a paramagnetic to ferromagnetic transition at $T$$_{c}$$\sim$125 K, accompanied by a semiconductor to metal transition [4]. The magnitude of CMR effect reaches 98$\%$ near $T$$_{c}$. Besides, this 327 bilayer manganese oxide exhibits a remarkable range of magnetic behavior, strongly depending on the concentration of Sr [9-14]. The substitution of the smaller Pr$^{3+}$ ions for the larger La$^{3+}$ ions causes a local lattice deformation: Jahn-Teller distortion (JT). In this system, MnO$_{6}$ octahedral distortions are interpreted in terms of the occupation ratio between $e$$_{g}$ and $t$$_{2g}$ level, and ordering of $e$$_{g}$ orbitals and their relationship to magnetic ground state [15]. The MnO$_{6}$ octahedra shrinks along the $c$-axis in the ground state with increasing $x$, implying an increase of the $d$($x$$^{2}$-$y$$^{2}$) orbital polarization [5-8]. On the other hand, the MnO$_{6}$ octahedra elongate along the $c$-axis in the ground state with increasing a Pr concentration $z$ at the Sr concentration of 0.4, implying an increase of the $d$(3$z$$^{2}$-$r$$^{2}$) orbital polarization. Furthermore, an increase of the Pr concentration causes a suppression of the planar ferromagnetic interaction, leading to a lowering of $T$$_{c}$. The ferromagnetic phase seems to vanish around $z$=0.5. This concentration area seems to be nearby the magnetically frustrated point (FP) in which ferro and antiferromagnetic interactions compete considerably with one another (although this origin is not clear yet). Since several vital interactions compete with each other, that is, the resultant instability of magnetic, lattice and band properties, it is expected that a unique phase or phenomena may be induced by external variables such as pressure and magnetic fields.

The ultrasonic measurement is a strong probe to examine the characteristic features of the 3$d$-electronic states of the Mn ions. The anomalies of the elastic constants is expected to be due to the coupling between magnetic moments or the conduction electrons and elastic strains associated with sound wave [19]. In this paper we present our investigations on the elastic and magnetic properties of single crystal LPSMO with the concentration of z=0.6. The preliminary reports have been published in Ref. [20-21].

\section{\label{sec:level1}Experiment}

A single crystal of LPSMO for $z$=0.6 was grown by the floating-zone (FZ) method using a mirror furnace. The specimen used in this study measures 5$\times$4 mm$^{2}$ in the $ab$-plane and 1 mm along the $c$-axis. The sound velocity ($v$) was measured by an ultrasonic apparatus based on a phase comparison method in magnetic fields up to 12 T generated by a superconducting magnet. Plates of LiNbO$_{3}$ were used for the piezoelectric transducers. The transducers were glued on to the parallel planes of the sample by an elastic polymer Thiokol. The absolute value of the elastic constant $C$=$\rho$$v$$^{2}$ by using the density $\rho$ of the crystal could not be estimated explicitly because the observed echo trains were not enough clear to do that. The high-field magnetization ($M$) measurement was carried out by using the vibrating sample magnetometer (VSM) method in fields up to 12 T and at temperatures down to 4.2 K. The detailed magnetization measurement was performed by superconducting quantum interference device (SQUID) magnetometer from 1.8 K to 300 K in magnetic fields up to 5 T.

\begin{figure}[h]
\begin{center}\leavevmode
\includegraphics[width=0.8\linewidth]{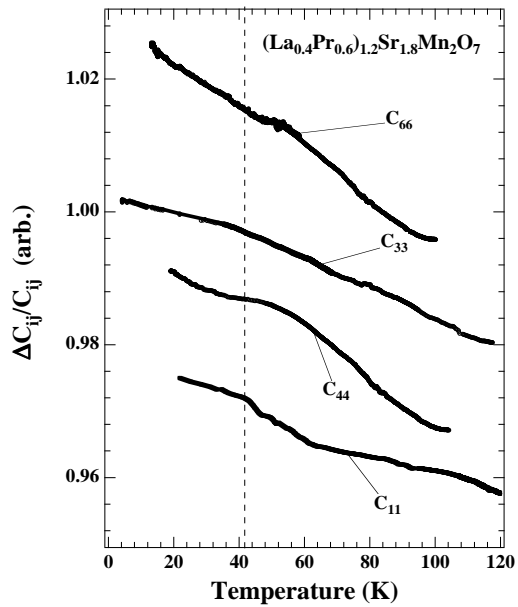}
\caption{Temperature dependence of elastic constants $C$$_{11}$, $C$$_{33}$, $C$$_{44}$ and $C$$_{66}$ of LPSMO for $z$=0.6 in the absence of magnetic fields. The dotted vertical line indicates a transition point determined by the resistivity measurement.}\label{figurename}\end{center}\end{figure}

\begin{figure}[h]
\begin{center}\leavevmode
\includegraphics[width=0.8\linewidth]{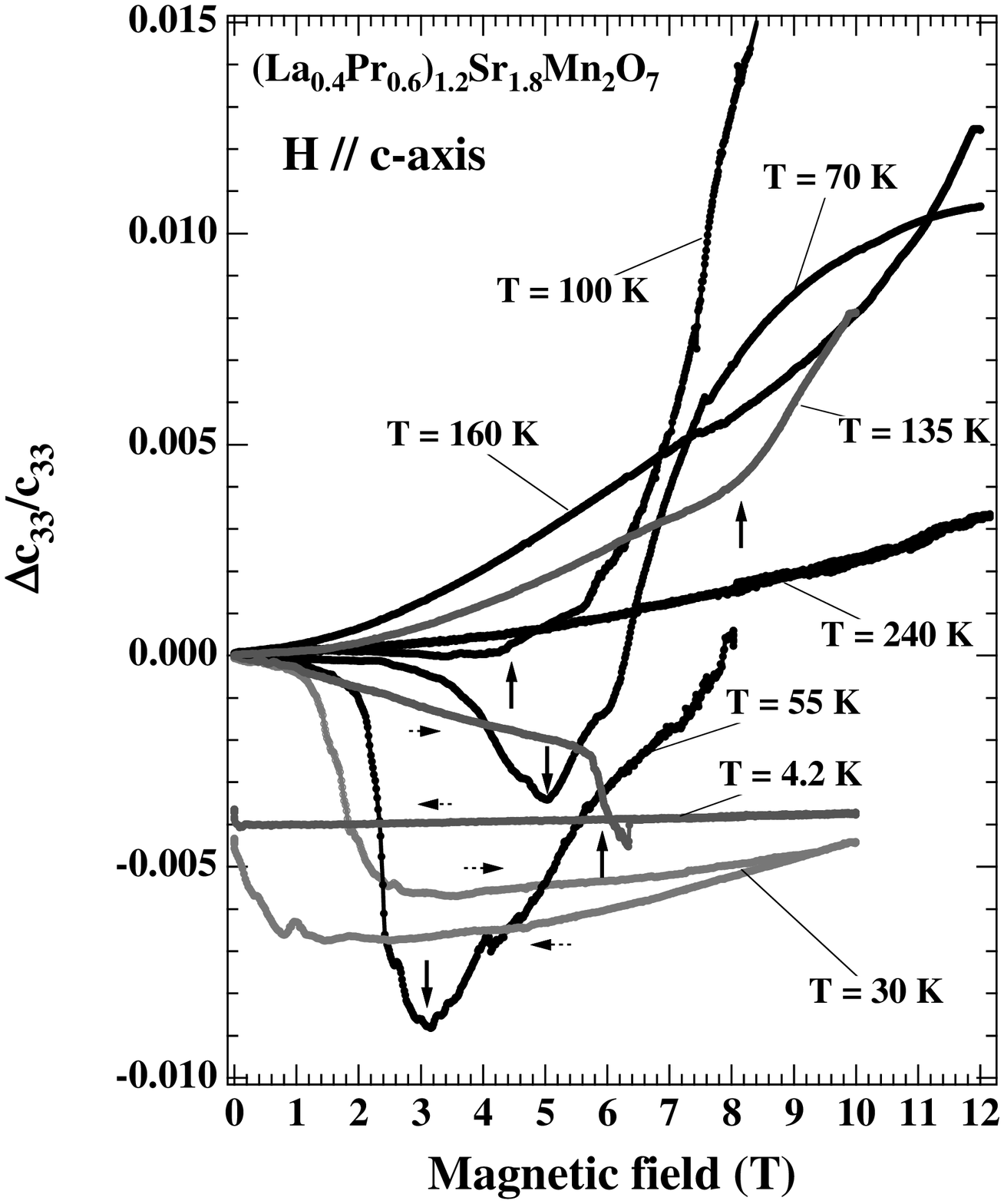}
\caption{Magnetic field dependence of elastic constant $C$$_{33}$ of LPSMO for $z$=0.6 at selected temperatures. The vertical arrows indicate a transition point. The horizontal ones indicate a direction of the applied fields.}\label{figurename}\end{center}\end{figure}

\begin{figure}[h]
\begin{center}\leavevmode
\includegraphics[width=0.8\linewidth]{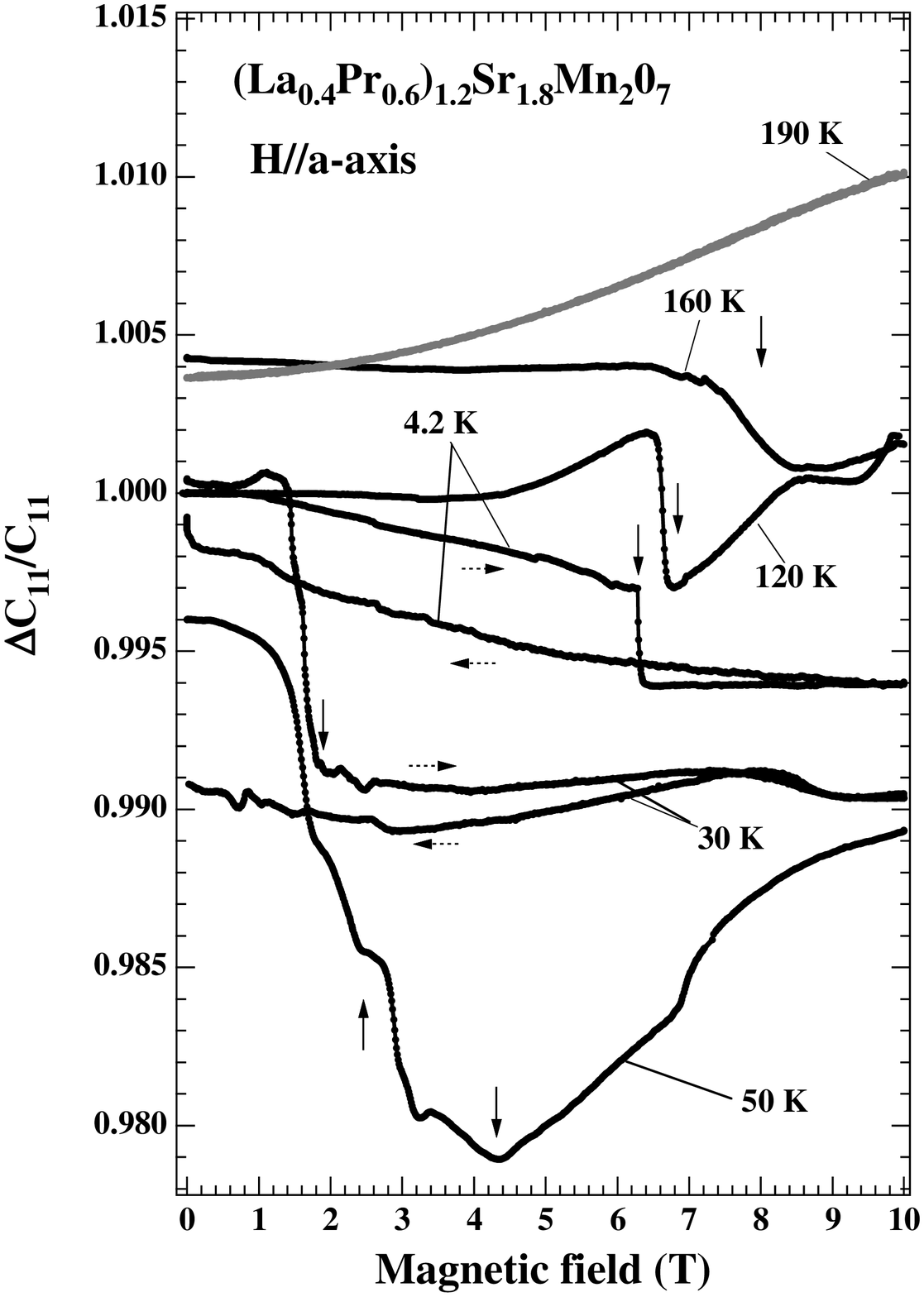}
\caption{Magnetic field dependence of elastic constant $C$$_{11}$ of LPSMO for $z$=0.6 at selected temperatures. The vertical arrows indicate a transition point. The horizontal ones indicate a direction of the applied fields.}\label{figurename}\end{center}\end{figure}

\begin{figure}[h]
\begin{center}\leavevmode
\includegraphics[width=0.8\linewidth]{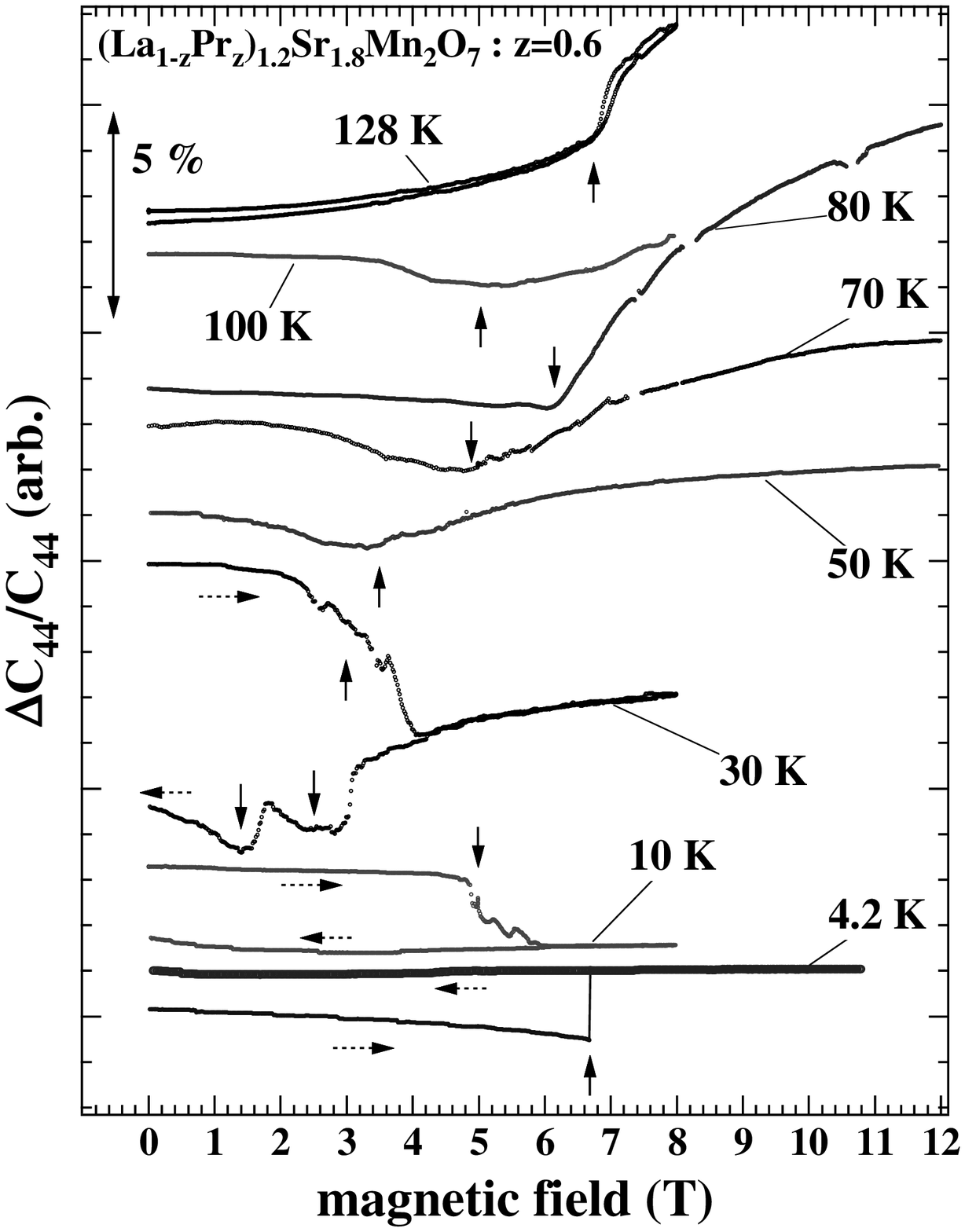}
\caption{Magnetic field dependence of elastic constant $C$$_{44}$ of LPSMO for $z$=0.6 at selected temperatures. The vertical arrows indicate a transition point. The horizontal ones indicate a direction of the applied fields.}\label{figurename}\end{center}\end{figure}

\begin{figure}[h]
\begin{center}\leavevmode
\includegraphics[width=0.8\linewidth]{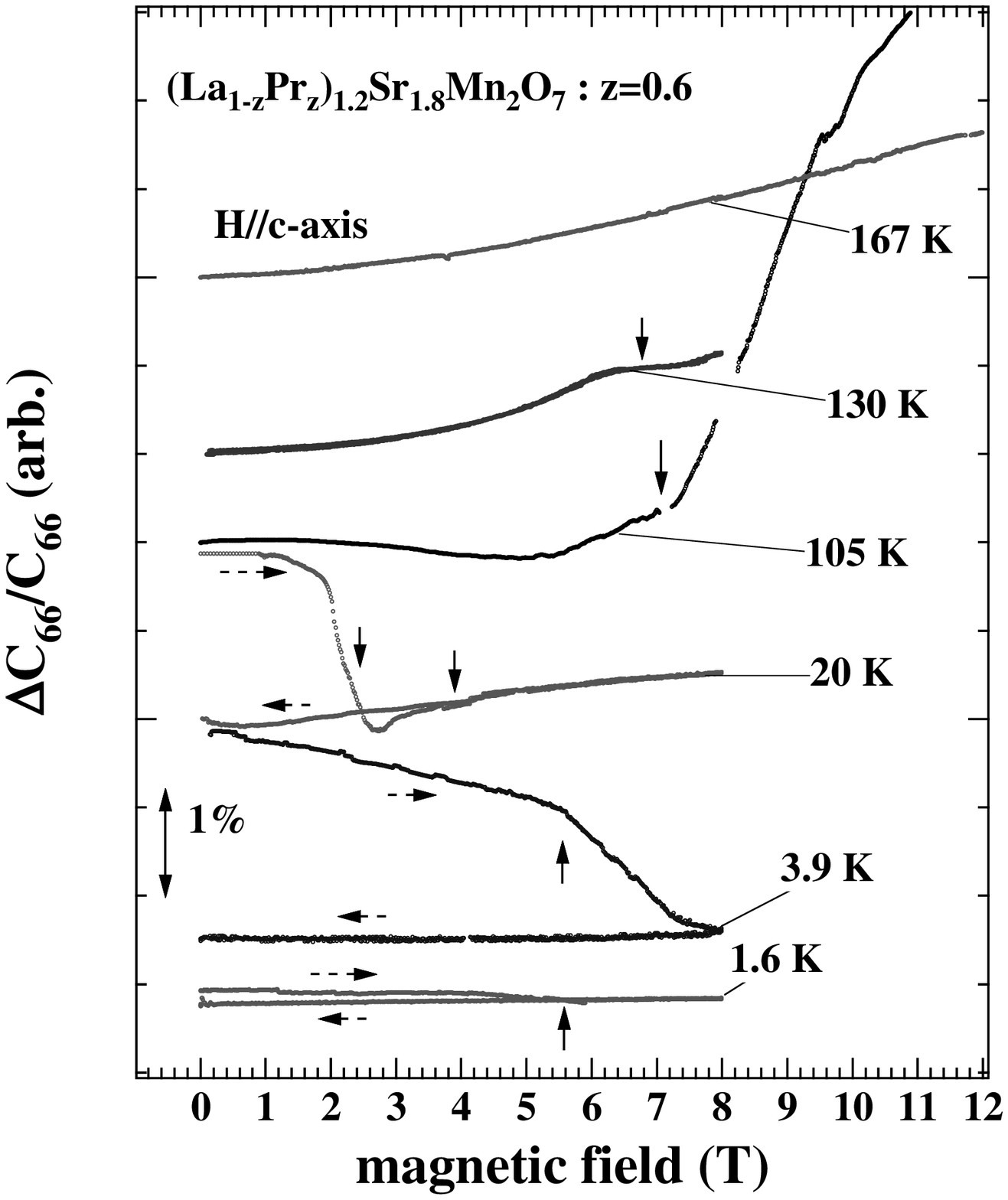}
\caption{Magnetic field dependence of elastic constant $C$$_{66}$ of LPSMO for $z$=0.6 at selected temperatures. The vertical arrows indicate a transition point. The horizontal ones indicate a direction of the applied fields.}\label{figurename}\end{center}\end{figure}

\section{\label{sec:level1}Experimental results}
\subsection{\label{sec:level2}Ultrasonic measurement}

In this section we report the elastic properties of LPSMO for $z$=0.6. Figure 1 shows the temperature dependence of the longitudinal elastic constant $C$$_{11}$, $C$$_{33}$ and the transverse elastic constant $C$$_{44}$, $C$$_{66}$ in the absence of a magnetic field. Plotted is the relative elastic constant change. To avoid overlap and complication of curves, their offsets are shifted. The same modification is utilized in all following figures in this section. $C$$_{11}$ and $C$$_{33}$ are the elastic constant determined by the longitudinal sound wave propagated along $a$-axis and $c$-axis, respectively, and $C$$_{44}$ and $C$$_{66}$ are those determined by the transverse one propagated along $a$-axis and $c$-axis, respectively. They both increase monotonically with decreasing a temperature. A slight anomaly seems to be observed around $T$$^{*}$ of 40 K in all elastic constants, where the temperature dependence of the resistivity exhibits a distinct shoulder structure[5-6]. A pronounced elastic anomaly, however, was induced by applying magnetic fields. Figure 2 shows the field dependence of $C$$_{33}$ at selected temperatures. At 4.2 K $C$$_{33}$ decreases slightly with increasing a field. The sharp decrease was observed at 6 T, and then $C$$_{33}$ increases slightly with the further increase of fields. Whereas, when the field decreases, $C$$_{33}$ decreases slightly with decreasing a field. On the contrary, no anomaly was observed at 6 T with decreasing a field. That is to say, the transition induced by a field is accompanied by a large hysteresis, indicating the transition to be first-order transition nature. The trace of $C$$_{33}$ is guided by the dotted arrows in Fig. 2. The transition field shifts to the lower temperature with increasing a field below around 40 K. Besides, the hysteresis turns to be smaller with increasing a field. Whereas, the transition field shifts to the higher temperature with increasing a temperature above around 40 K. Furthermore, $C$$_{33}$ increases abruptly above the transition field in the higher fields above around 40 K. It is noted that no hysteresis accompanied by the transition was observed above 40 K. $C$$_{33}$ slightly increases with increasing a field and no anomaly was recognized at 240 K. Figure 3 shows the field dependence of $C$$_{11}$ at selected temperatures. The general behavior is similar to that of $C$$_{33}$. However, the elastic anomaly observed at 120 K and 160 K is opposite to that observed for $C$$_{33}$. The $C$$_{11}$ decreases rapidly at the transition field.

 Figures 4 and 5 show the field dependence of $C$$_{44}$ and $C$$_{66}$ at selected temperatures, respectively. It is noted that $C$$_{44}$ exhibits a sharp jump at the transition field at 4.2 K, in contrast to that of $C$$_{11}$ and $C$$_{33}$. It is accompanied by a distinct hysteresis as well. On the other hand, $C$$_{66}$ exhibits a slight drop at the transition field at 1.6 K. The general behavior around the transition field and the elastic constants are qualitatively similar to those of the longitudinal $C$$_{11}$ and $C$$_{33}$. A distinct transition accompanied by a large hysteresis was observed at around 5 T in the field dependence of elastic constants at 4.2 K. The transition shifts to lower temperatures and the accompanied hysteresis becomes smaller with increasing temperature up to about 50 K. On the other hand, the transition shifts to higher temperatures with the further increase in temperature. This transition is not accompanied by a hysteresis within the experimental error.

\begin{figure}[h]
\begin{center}\leavevmode
\includegraphics[width=0.8\linewidth]{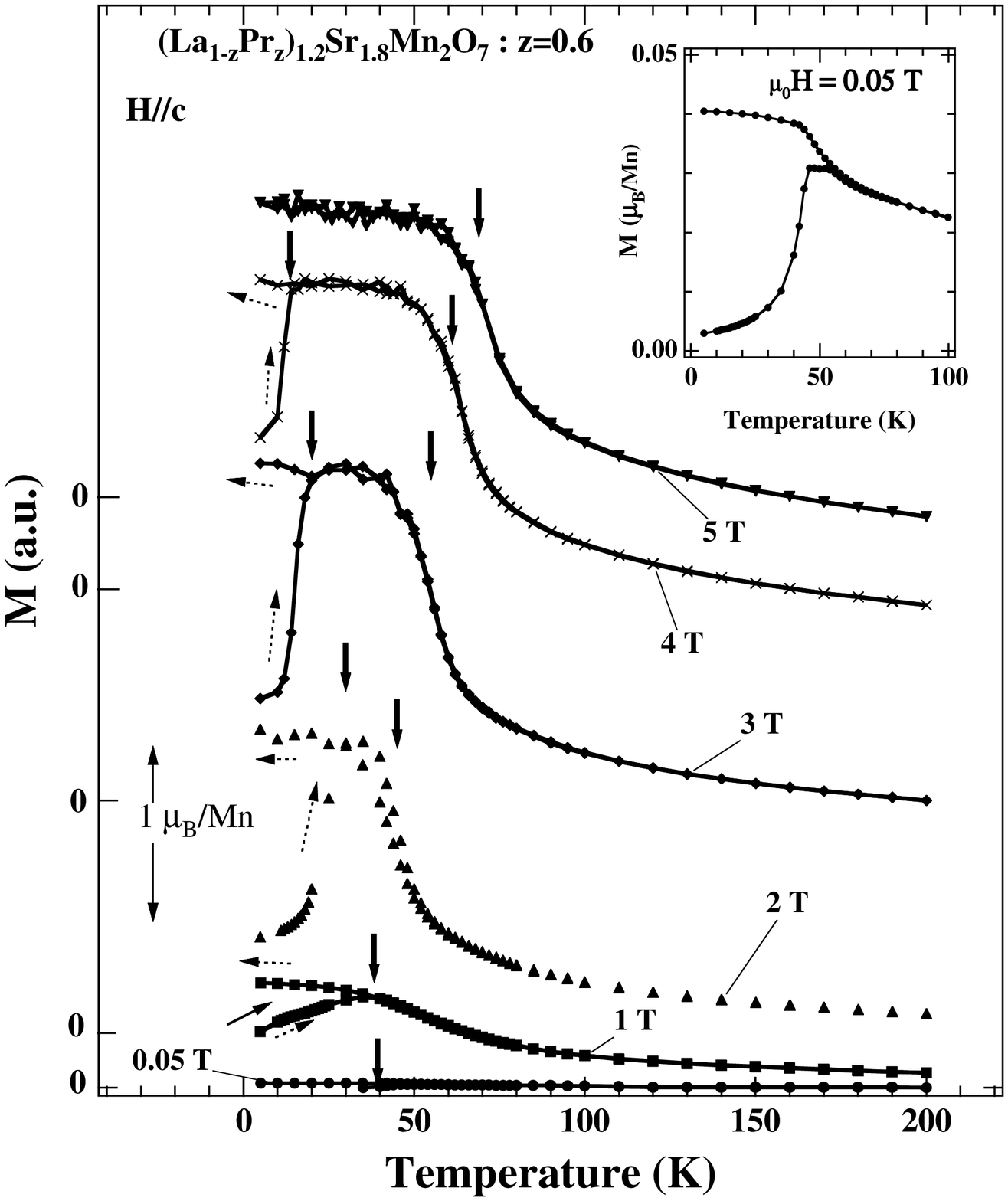}
\caption{Temperature dependence of the magnetization in selected fields below 5 T along $c$-axis. Inset shows the behavior around 40 K in a field of 0.05 T on an expanded scale. The solid arrows indicate a transition point. The dotted arrows indicate a direction of the applied fields.}\label{figurename}\end{center}\end{figure}

\begin{figure}[h]
\begin{center}\leavevmode
\includegraphics[width=0.8\linewidth]{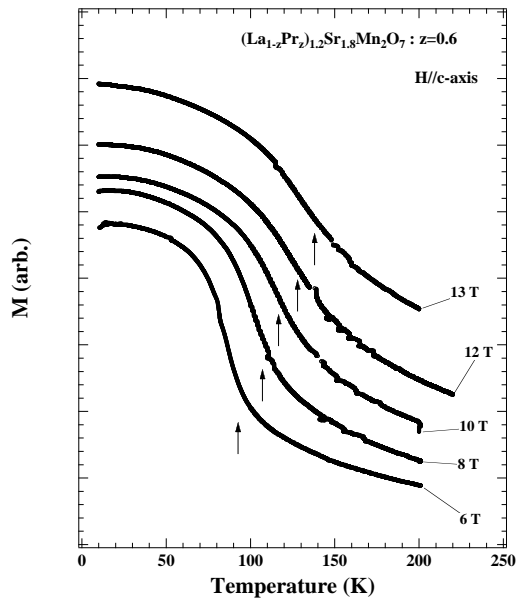}
\caption{Temperature dependence of the magnetization in selected fields above 5 T along $c$-axis. The arrows indicate a transition point.}\label{figurename}\end{center}\end{figure}

\begin{figure}[h]
\begin{center}\leavevmode
\includegraphics[width=0.7\linewidth]{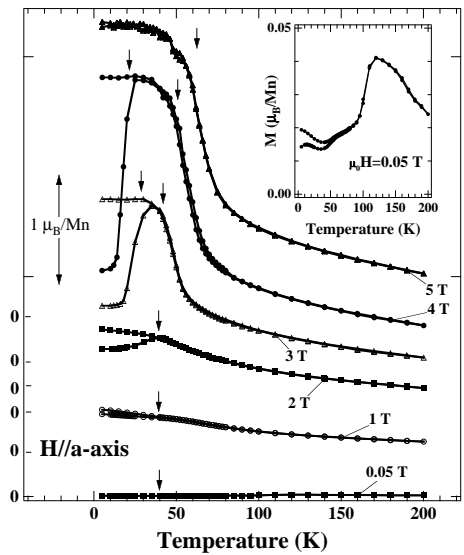}
\caption{Temperature dependence of the magnetization in selected fields below 5 T along $a$-axis. The arrows indicate a transition point. Inset shows the behavior around 40 K in a field of 0.05 T on an expanded scale. }\label{figurename}\end{center}\end{figure}

\begin{figure}[h]
\begin{center}\leavevmode
\includegraphics[width=0.8\linewidth]{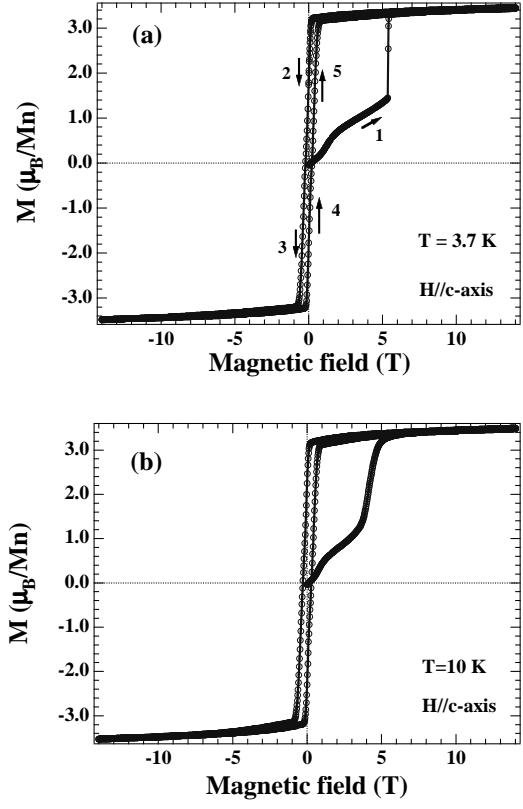}
\caption{Field dependence of the ZFC magnetization for the field along $c$-axis at 3.7 K (a) and 10 K. (b)}\label{figurename}\end{center}\end{figure}

\begin{figure}[h]
\begin{center}\leavevmode
\includegraphics[width=0.8\linewidth]{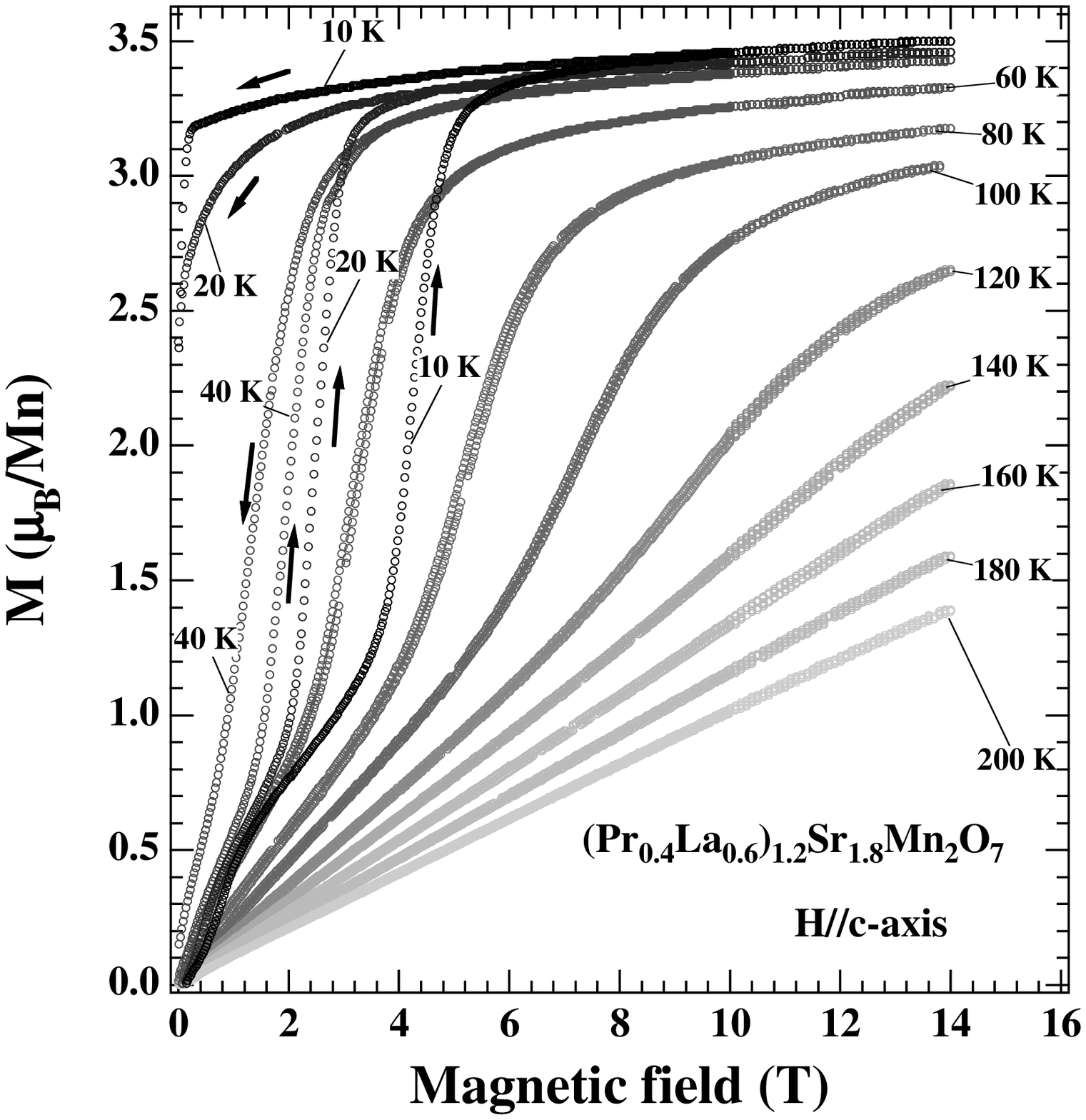}
\caption{Field dependence of the magnetization for the field along $c$-axis above 20 K. }\label{figurename}\end{center}\end{figure}

\begin{figure}[h]
\begin{center}\leavevmode
\includegraphics[width=0.8\linewidth]{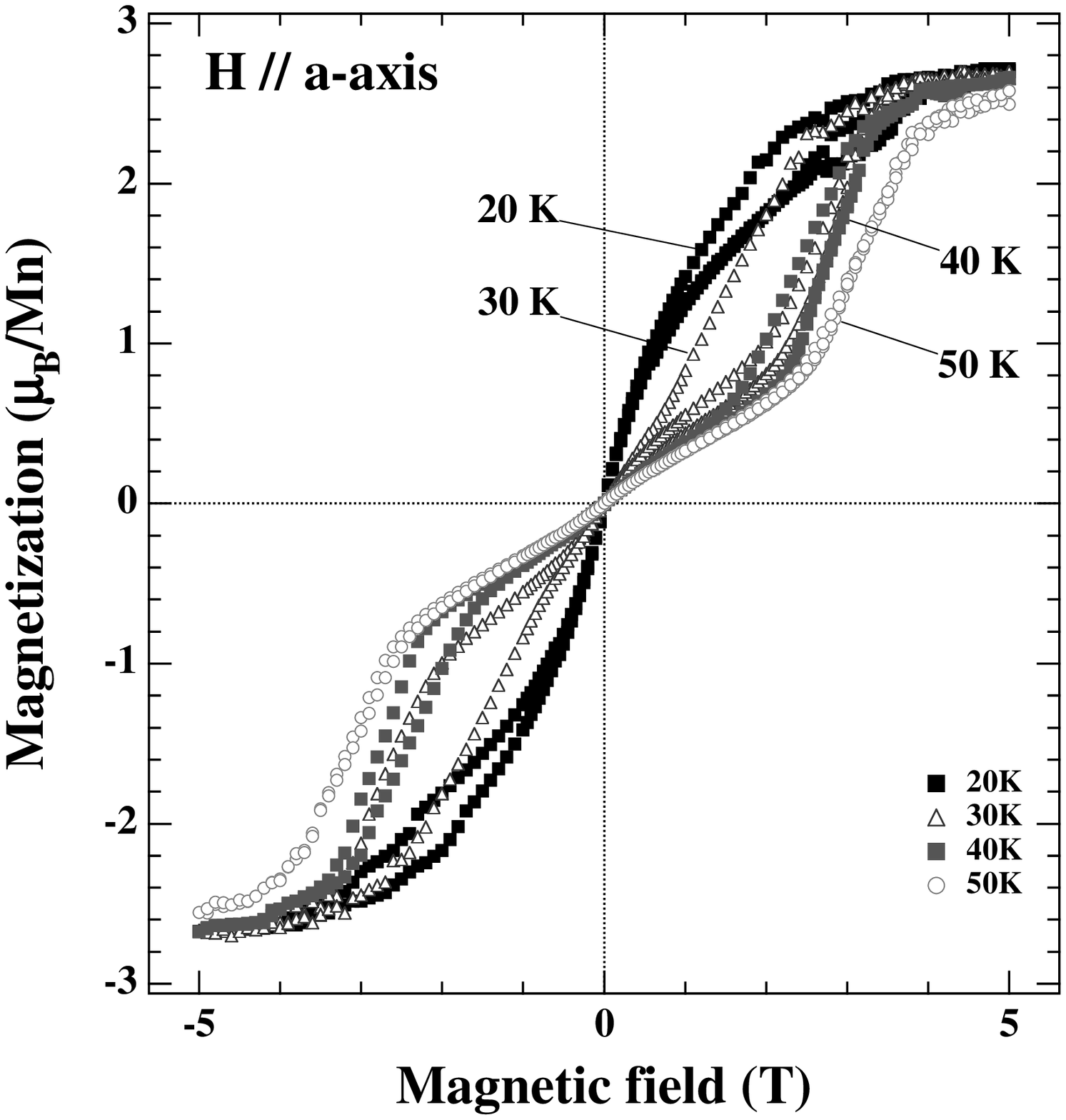}
\caption{Field dependence of the magnetization for the field along $a$-axis above 20 K. }\label{figurename}\end{center}\end{figure}

\begin{figure}[h]
\begin{center}\leavevmode
\includegraphics[width=0.8\linewidth]{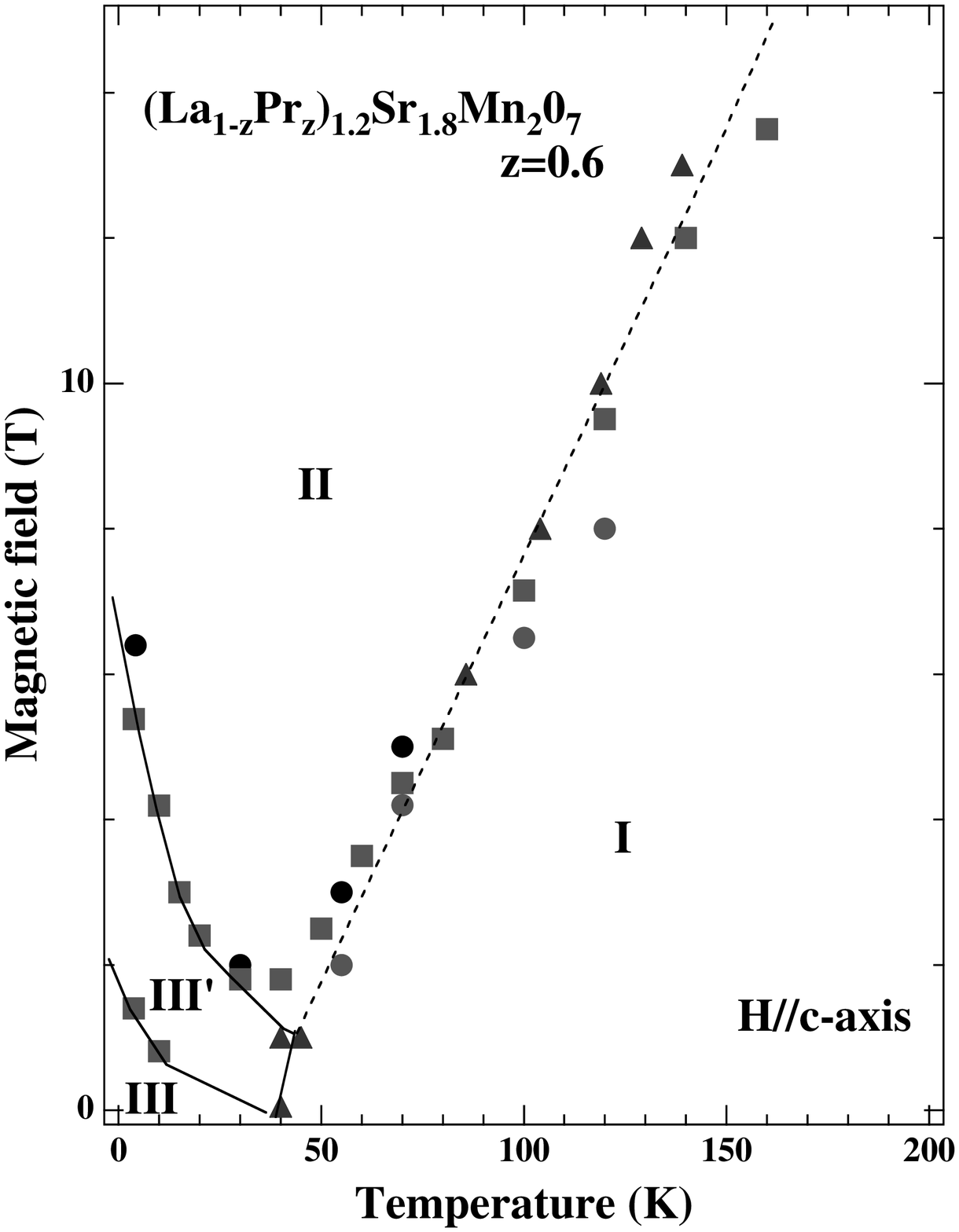}
\caption{($H$-$T$) magnetic phase diagram determined by the present results. The squares, triangles and circles denote the transition points determined by the $M$-$H$ curves, $M$-$T$ curves and elastic constant measurement, respectively. The solid and dotted lines are a guide to the eye. The solid and dotted lines denote the first- and second- nature phase transition, respectively}\label{figurename}\end{center}\end{figure}

\subsection{\label{sec:level2}Magnetization measurement}
In this section we report the results of the magnetization measurement of LPSMO for $z$=0.6. Figure 6 shows the temperature ($T$) dependence of the magnetization in selected fields along $c$-axis $M$$_{c}$. $M$$_{c}$-$T$ curve exhibits a slight increase without a distinct anomaly with decreasing temperature in a field of 0.05 T. Inset shows the behavior around 40 K on an expanded scale. A clear anomaly is recognized, although it is fairly small. A distinct cusp, however, appears in the zero-field-cooled (ZFC) magnetization curve at around 40 K in a field of 1 T. On the other hand, the field-cooled (FC) magnetization curve exhibits a gradual increase below about 100 K and reaches a constant value in the further decrease in temperature, without any subsequent drop below 40 K. It is noted that the FC magnetization curve is quite different from ZFC one below around 40 K. An increase of magnetization below 40 K becomes steeper and a hysteresis becomes larger with increasing a field. Furthermore, the anomaly at around 40 K under a field of 1 T split into two anomalies. The lower transition temperature shifts to lower temperatures up to 4 T, accompanied by a distinct hysteresis and disappears around 5 T. While, the higher transition temperature shifts to higher temperature. It is noteworthy that the higher transition is not accompanied by a hysteresis, indicating a second-order transition nature. This transition becomes steeper up to 6 T, and then smeared with the further increase of field as shown in Fig. 7. These transitions correspond to those observed in the elastic constants, consistently. Fig. 8 shows the temperature ($T$) dependence of the magnetization in selected fields along $a$-axis $M$$_{ab}$. Inset shows the behavior around 40 K on an expanded scale. Similar to those of $M$$_{c}$-$T$ curve, a clear anomaly is recognized, although it is fairly small. A distinct cusp, however, appears in the zero-field-cooled (ZFC) magnetization curve at around 40 K in a field of 1 T. The similar behaviors such as hysteresis, a split of transition and a steep increase of magnetization were observed under the higher fields. Here, let us compare the FC magnetization curve in a field of 0.05 T between along $c$-axis $M$$_{c}$ and in $a$$b$-plane $M$$_{ab}$ as shown in inset of Figs. 6 and 8, respectively. One can see that a magnetic moment lies in $a$$b$-plane by their anisotoropic behavior.

 Next, we show the high-field magnetization process, $M$-$H$ curve. The virgin $M$$_{c}$-$H$ curve at 3.7 K exhibits a steep increase around 5 T as shown in Fig. 9 (a). A small shoulder was also observed around 2 T only in the virgin process with respect to an increase of fields. A clear rectangular hysteresis loop was observed in the field range between around 1 T and -1 T. The saturated magnetization reaches 3.5 $\mu$$_{B}$/Mn. This hysteresis loop becomes smaller with increasing temperatures as shown in Fig. 9 (b). Figure 10 shows the high-field magnetization measurement above 10 K. A metamagnetic transition observed in the virgin $M$$_{c}$-$H$ curve shifts to lower fields with increasing temperatures up to 50 K. While, the transition shifts to higher fields and becomes gentler with the further increase of temperature. It is noted that the transition is accompanied by no hysteresis above 50 K. No metamagnetic transition was observed at 200 K. Figure 11 shows the $M$$_{ab}$-$H$ curve at selected temperatures. Similarly, a clear step increase was observed in $M$$_{ab}$-$H$ curve as well as $M$$_{c}$-$H$ one. The transition field shifts to higher fields and the accompanied hysteresis becomes smaller with increasing a temperatures. The saturated magnetization reaches 2.5 $\mu$$_{B}$/Mn, which is almost 70$\%$ of $M$$_{c}$, indicating that the $c$-axis is easy axis of magnetization.

\subsection{\label{sec:level2}Magnetic phase diagram}
 The magnetic ($H$-$T$) phase diagram was made by the measured elastic constants and magnetizations for the field along $c$-axis, i.e. easy axis of magnetization. Figure 12 shows the ($H$-$T$) phase diagram under the field along $c$-axis. There are mainly three different regions. The phase I and II are considered to be paramagnetic and Induced ferromagnetic phase, respectively. It is deduced from that almost full magnetic moment of 3.5 $\mu$$_{B}$/Mn is induced in the phase II, where the system exhibits a metallic behavior in the resistivity. The phase III is the field-induced ferromagnetic phase. There are two sub-regions in the phase III. The boundary is defined well only by the anomaly in the virgin $M$-$H$ and $C$-$H$ curves. This boundary may be due to the domain rotation. It is noted that the boundary between I and III under zero field is quite obscure. However, it becomes distinct in a magnetic field, indicating that the phase III appears only in a magnetic field. The transition is accompanied by a remarkable hysteresis in both magnetization and elastic constants at the boundary between the phase III and II, being indicative of a first-order transition nature. On the contrary, no hysteresis is accompanied at the boundary between the phase I and II, being indicative of a second-order transition nature. The obtained qualitative features of elastic constants are summarized as follows. The elastic constant decreases crossing the phase boundary from the phase III to II accompanied by a remarkable hysteresis only in the virgin process with increasing a field. Whereas, it decreases gradually with increasing a field in the phase I, and increases gradually after crossing the boundary from the phase I to II without any hysteresis. The temperature of around $T$$^{*}$ of 40 K is likely to be the critical temperature to distinguish these phases, although $T$$^{*}$ is quite vague.

\begin{figure}[h]
\begin{center}\leavevmode
\includegraphics[width=0.8\linewidth]{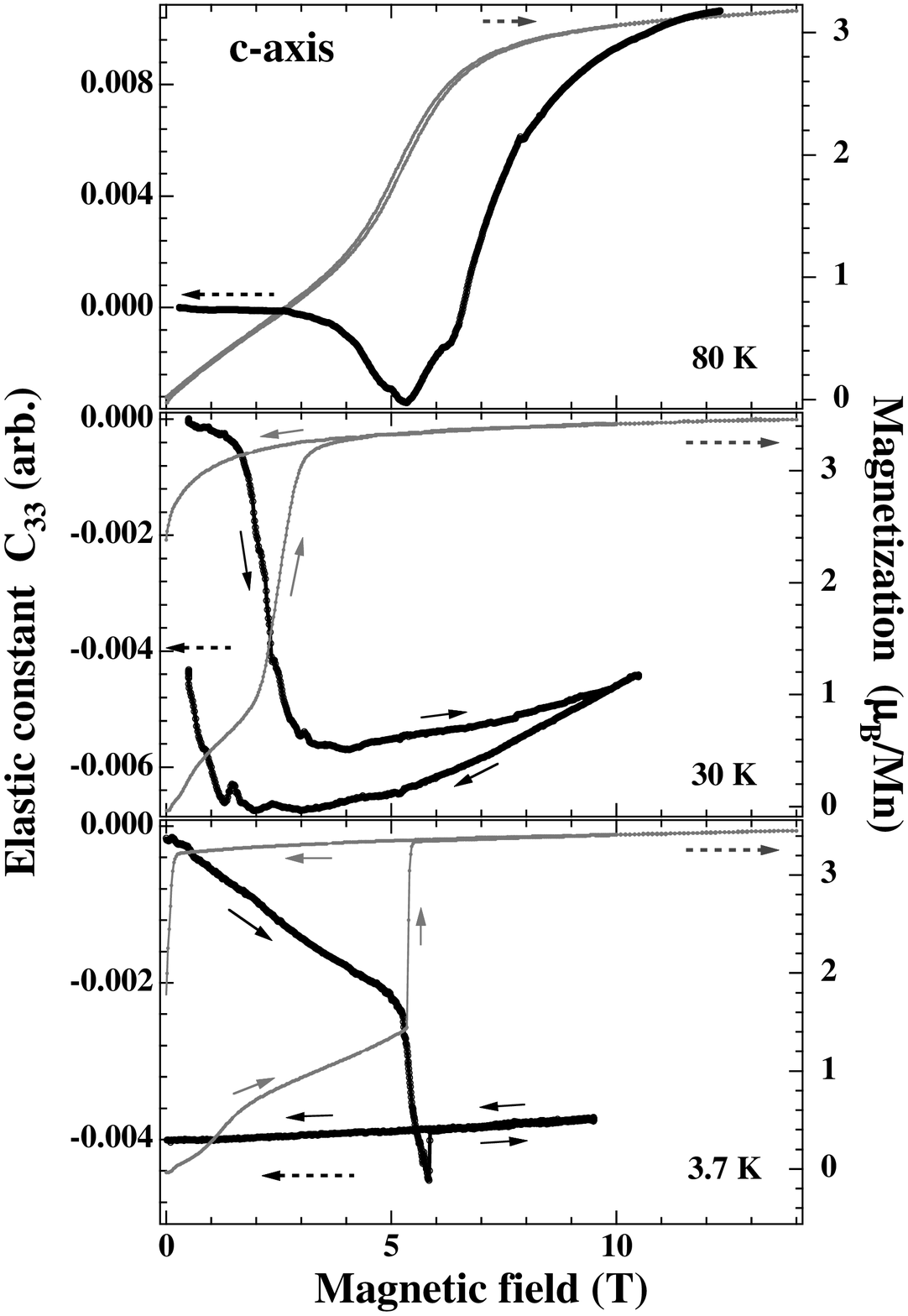}
\caption{Comparison of the obtained elastic constants with the magnetization at the temperature of 3.7 K, 30 K and 80 K. }\label{figurename}\end{center}\end{figure}

\begin{figure}[h]
\begin{center}\leavevmode
\includegraphics[width=0.8\linewidth]{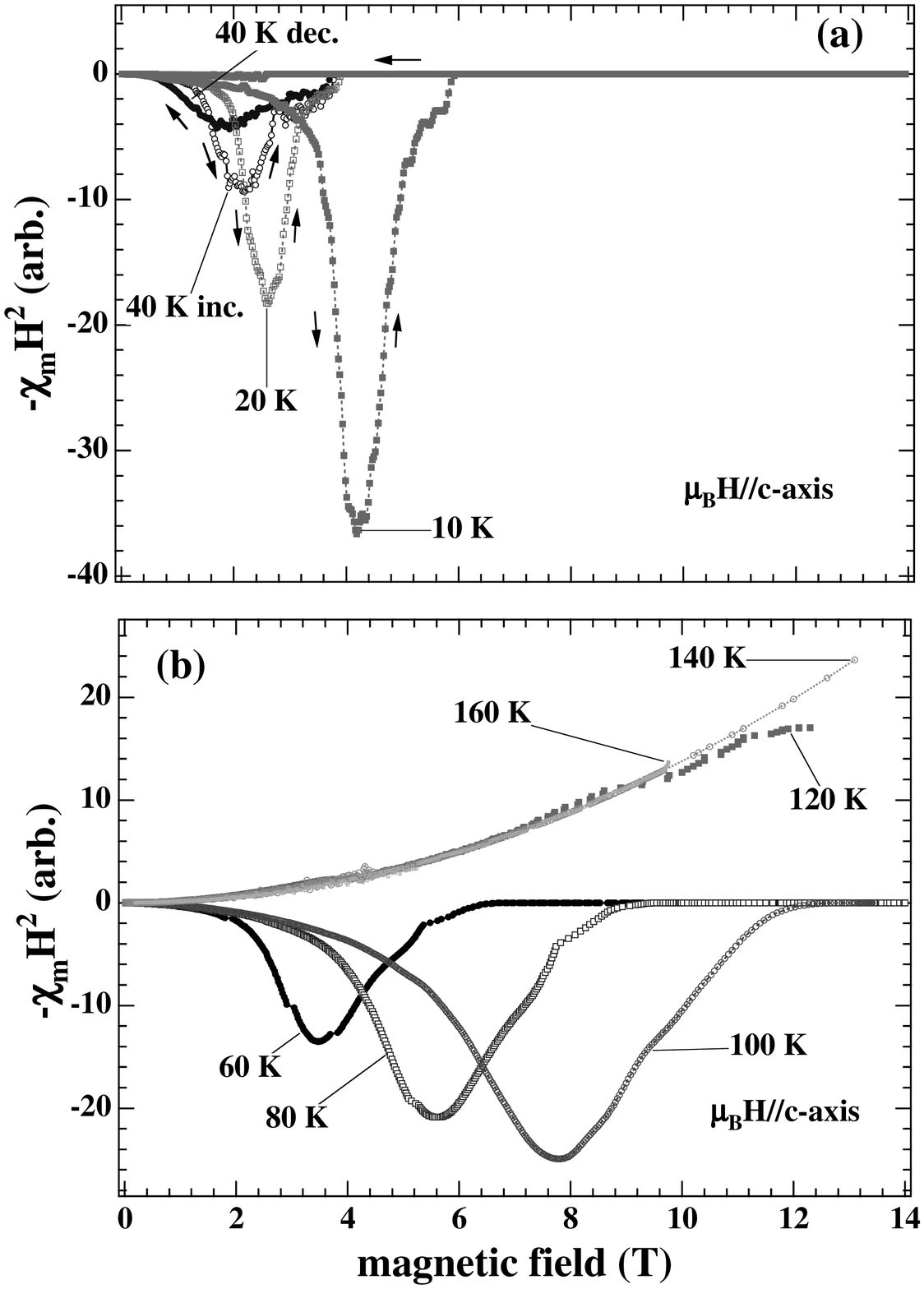}
\caption{The field dependence of -$H$$^{2}$$\chi$$_{m}$ (a) below the critical temperature of 40 K and (b) above that, respectively. The dotted arrows indicate a direction of the applied fields. 40 K inc. and 40 K dec. denotes a result at 40 K in the increase and decrease process with respect to a magnetic field, respectively. }\label{figurename}\end{center}\end{figure}

\section{\label{sec:level1}Discussions}
 First, we discuss the obtained elastic anomalies in the field dependence of elastic constants. Let us compare them with the magnetization curves as shown in Fig. 13. A sharp softening accompanied by a prominent hysteresis was observed at metamagnetic transition field $H$$_{c}$ at $T$= 3.7 K and 30 K. On the other hand, a softening without any hysteresis was observed when crossing the transition field at the temperature of 80 K. The critical temperature is roughly estimated to be 40 K. From this result, one can see the correlative relation between them. It is likely that the elastic anomalies accompanied with a metamagnetic transition are responsible for electron-phonon effect. In such a case the elastic softening is related to the magnetic susceptibility, $\chi$$_{m}$=$\partial$$M$/$\partial$$H$[22-24]. A strong coupling between elastic strains associated by sound waves and relevant conduction bands is usually described in terms of the magnetic Gr$\ddot{u}$neisen parameter $\Gamma$$_{B}$ = -$\partial$ln$H$$_{c}$/$\partial$$\varepsilon$$_{\Gamma}$, where $\varepsilon$$_{\Gamma}$ is the elastic strain with $\Gamma$ symmetry. Then, the elastic softening is given by 
\begin{eqnarray}
\Delta C_{\Gamma}=- \Gamma _{B}^{2}H^{2}\chi_{m}
\end{eqnarray}
Figure 14 (a) and (b) show the field dependence of -$\chi$$_{m}$$H$$^{2}$ below the critical temperature $T$$^{*}$ of 40 K and above that, respectively. As compared to the measured field dependence of elastic constants, especially longitudinal ones shown in Figs. 2 and 3, a good agreement with -$\chi$$_{m}$$H$$^{2}$ is achieved by adopting the single free parameter $\Gamma$$_{B}$ at various temperatures, described by eq. (1). Especially in the case above 40 K, we can obtain a good agreement, although some of them in the high field region and their quantitative aspects do not agree well between them. We emphasize that this agreement gives evidence for a strong spin-lattice coupling in LPSMO below $T$$^{*}$ of 40 K. The expected upturn of elastic constant by -$\chi$$_{m}$$H$$^{2}$ above $H$$_{c}$ below 40 K doesn't appear in the measured elastic constants, which is expected by this calculated model. One scenario to explain the discrepancy is that the change of carrier number probably causes the difference as discussed below. 

Next, let us discuss the coupling between the elastic strains and the conduction electrons. According to the deformation potential coupling theory, the coupling can be described as 
\begin{eqnarray}
E_{k}(\varepsilon)= E_{k}^{0} + d_{\Gamma}(k) \varepsilon _{\Gamma}
\end{eqnarray}
where, $E$$_{k}$ denotes a single particle electron state and $d$$_{\Gamma}$ the deformation potential coupling constant with $\Gamma$ symmetry [25-27]. The suffix $\Gamma$ denotes the irreducible representation in the tetragonal symmetry of the present system [19]. The free energy for conduction electrons $F$$_{el}$ may be expressed as follows,
\begin{eqnarray}
F_{el} = nE_{F} - k_{B}T\sum_{k}ln(1 + exp((E_{F} - E_{k})/k_{B}T)
\end{eqnarray}
where $E$$_{F}$ and $n$ denote the Fermi energy and the number of the conduction electrons in the band, respectively. From the second derivative of the free energy, one obtains the expression of a relative change in the elastic constant.
\begin{eqnarray}
\Delta C = - \int \rho (\varepsilon) \frac{\partial f_{k}}{\partial \varepsilon } d\varepsilon
\end{eqnarray}
where, $\rho$($\varepsilon$) and $f$$_{k}$ denote the density of state and Fermi-Dirac distribution function, respectively. For simplicity the coupling constant $d$$_{\Gamma}$ is taken to be independent of $k$ (rigid band model), in a two band model in which we assume them to be $d$($x$$^{2}$-$y$$^{2}$) and $d$(3$z$$^{2}$-$r$$^{2}$). Then, the above formula leads to the "band-Jahn-Teller effect" one. It is expected that an increase of the density of states, i.e., the carrier number contributes greatly to $\Delta$$C$. Thus, a decrease of the resistivity as a function of magnetic fields leads to a decrease of the elastic constant as described by eq. (4). We conjecture that this masks partly the expected upturn just above $H$$_{c}$. Unfortunately, we cannot discuss these changes quantitatively at present since the absolute values of the elastic constants were not estimated. Nevertheless, we suggest that the elastic anomaly and the absence of the upturn of elastic constant above $H$$_{c}$ below $T$$^{*}$ are to be explained by the coupling between elastic strain and $\chi$$_{m}$, and between elastic strain and carriers, respectively.

Secondly we would like to discuss the phase III. The obtained features may suggest this phase to be fairly metastable phase. Figs. 2, 3 and 4 give typical examples to describe this situation. Actually, the oscillatory phenomena is observed for $C$$_{33}$ at 30 K, $C$$_{11}$ at 30 K and 50 K and $C$$_{44}$ at 10 and 30 K below $H$$_{c}$. The oscillations disappear above $H$$_{c}$ and for higher temperatures. We conjecture that these oscillations may be related to the vicinity of magnetic instability, such as a spin-glass state, since the transition is accompanied by a prominent hysteresis in both results of ultrasonic and magnetization measurements. However, it is somehow unusual, because a slight anomaly is observed in the absence of a magnetic field. A spin glass-like behavior shows up only in a magnetic field. The critical temperature in the absence of a field is estimated to be around $T$$^{*}$ of 40 K by interpolation of the boundary between the phase I and III. Again, this temperature corresponds to that at which a bend is observed in the resistivity curve. A slight anomaly was observed in both the temperature dependence of elastic constant and magnetization curve at this temperature. (La$_{1-z}$Nd$_{z}$)$_{1.2}$Sr$_{1.8}$Mn$_{2}$O$_{7}$ and (La$_{0.8}$Gd$_{0.2}$)$_{1.4}$Sr$_{1.6}$Mn$_{2}$O$_{7}$ would be helpful to discuss this phase, which was reported previously by Moritomo $et$ $al$ [28] and Dho $et$ $al$ [29], respectively. The similar phase shows up in (La$_{1-z}$Nd$_{z}$)$_{1.2}$Sr$_{1.8}$Mn$_{2}$O$_{7}$ ($z$=0.4) at low temperature. In comparison, this phase is presumably due to metastable region where the parainsulator and ferrometal state can coexist. That is to say, ferromagnetic clusters can be grown by applying a magnetic field. This phase was understood within the framework of spin degree of freedom so far. Our previous results of LPSMO for $z$=0.6 indicated that magnetostriction (MS) along the $c$-axis and in the $a$$b$ plane showed a sudden decrease and increase at the transition field, respectively [7-8]. Furthermore, it should be noted that the transition is accompanied by a prominent hysteresis in MS. This implies that an orbital degree of freedom plays a crucial role in this ordered phase III as well. The hysteresis region is observed in the elastic constants as a function of field becomes broader with decreasing a temperature in the phase III. Classical thermodynamics indicates that the first-order phase transition of the metastable state can occur under the condition that the potential barrier between two states is comparable with temperature [28]. Thus it becomes difficult to transit between the metastable states with decreasing a temperature. This energy difference may correspond to $T$$^{*}$ of 40 K. The results of elastic constants and MS as a function of fields imply that the two states: paramagnetic and ferromagnetic states have the each different state with respect to the $e$$_{g}$ state. As was seen in LPSMO ($z$=0) the $e$$_{g}$ state is dominated by the $d$(3$z$$^{2}$-$r$$^{2}$) in the ferromagnetic phase, whereas an amount of $d$($x$$^{2}$-$y$$^{2}$) population mixes into $d$(3$z$$^{2}$-$r$$^{2}$) component in the paramagnetic phase. Recently, this behavior was confirmed microscopically by the neutron scattering measurement [30-31]. We speculate that this situation may provide the spatial disorder of the orbital state in addition to the magnetic disorder.

Finally, we would like to comment on nature of phase II and III. A distinct hysteresis was observed in phase III. There is another boundary in phase III, distinguished by III and III'. It seemed reasonable to think that the boundary between phase III and III' may be due to the rotation of a magnetic domain wall. We expect that the orbital glass state with respect to $d$($x$$^{2}$-$y$$^{2}$) and $d$(3$z$$^{2}$-$r$$^{2}$) or/and orbital phase clusters: para- and ordered-cluster with $d$(3$z$$^{2}$-$r$$^{2}$) component could occur in the phase III because the two orbital states can coexist at each Mn sites in the phase III in the above-mentioned way. That is to say, it seems that the ferromagnetic cluster arises because the two levels are close to each other in energy, and the occupation ratio strongly affects on the magnetic property. This situation lets us conjecture that the occupation ratio of $d$(3$z$$^{2}$-$r$$^{2}$) are different at each Mn sites. As a result two kinds of the nearly-degenerate orbital: $d$($x$$^{2}$-$y$$^{2}$) or $d$(3$z$$^{2}$-$r$$^{2}$) could be arranged randomly at Mn sites, and cause the resultant spin-glass behavior. Furthermore, this metastable state could vanish in phase II, implying a uniform magnetic and a resultant uniform orbital state could be induced in phase II. The experimental fact that no hysteresis is observed in phase II supports this scenario. A rapid decrease of a relaxation time $\tau$ of the magnetization and magnetostriction in phase II also supports this uniform state with respect to magnetic and orbital properties, $i$.$e$., disappearance of the glass state: $metastable$ $state$ in phase II.[5] This is left to future studies. In order to confirm this point, the frequency dependence of sound wave on elastic constants is currently in progress.

\section{\label{sec:level1}Concluding remarks}
 We performed magnetization and ultrasonic measurements on LPSMO for $z$=0.6 single crystal as a function of temperature and magnetic field. A pronounced elastic anomaly was observed across the boundary of the each magnetic phases. We gave a useful phenomenological description of these results in terms of the coupling between elastic strain and $\chi$$_{m}$=$\partial$$M$/$\partial$$H$. The microscopic experimental and theoretical descriptions are still lacking. In addition, we found a new metamagnetic-like transition above $T$$^{*}$ of 40 K in the magnetization measurement accompanying no hysteresis, indicative of second-transition nature. ($H$-$T$) magnetic phase diagram was determined, and it was found that the field-induced magnetic phase appears. This field-induced phase has a remarkable hysteresis, being indicative of the strong coupling among the spin, orbital and lattice of Mn ion. It is noteworthy that a remarkable hysteresis in the elastic constants and also in magnetostrictions across the phase boundary of III was observed. A comparison between the field dependence of elastic constants and magnetization gives their close relation combined by a single parameter $\Gamma$$_{B}$, indicating a strong spin-lattice coupling in LPSMO. There still remains the fundamental question why the phase boundary around $T$$^{*}$ in the absence of field is not distinguished. For a further and qualitative discussion, it is necessary to make clear the magnetic structure in the ordered phase III and III', and a microscopic description in the ordered states.

\begin{acknowledgments}
We are grateful to Y. Tobinai, H. Uematsu and M. Nakamura for their help in the ultrasonic measurement and the operation of the cryogenic apparatus. The measurements have been performed in the Cryogenic Division of the Center for Instrumental Analysis, Iwate University. This work was partly supported by a Grant-in-Aid for Science Research from the Minister of Education, Culture, Sports, Science, and Technology of Japan.
\end{acknowledgments}

\end{document}